\documentclass[11pt,twoside]{article}
\usepackage{asp2004}
\begin{document}
\title{The evolution of massive stars: 
a selection of facts and questions
}
\author{Dany Vanbeveren}
\affil{Astrophysical Institute, Vrije Universiteit Brussel, Pleinlaan 2, 1050 Brussels, Belgium.\\
dvbevere@vub.ac.be
}

\begin{abstract}

\noindent In the present paper we discuss a selection of facts and questions related to observations and 
evolutionary calculations of massive single stars and massive stars in interacting binaries. We
focus on the surface chemical abundanceÕs, the role of stellar winds, the early Be-stars, the high
mass X-ray binaries, the effects of rotation on stellar evolution. Finally, we present an
unconventionally formed object scenario ({\it UFO-scenario}) of WR binaries in dense stellar
environments.

\end{abstract}
\thispagestyle{plain}

\section{Introduction}

The overall evolution of massive stars is critically affected by uncertainties in various physical 
processes which determine the stellar structure. To illustrate, we do not know whether
semi-convection is slow or fast, whether convective core overshooting is small or large. There is no
unanimousness about the effects of rotation on stellar evolution (see also section 6). It is a fact however
that for the most massive stars stellar wind mass loss is important in order to understand their evolution
(section 3). Moreover, the evolution of massive close binaries is determined by processes which are a fact
but where the physics is in some cases only a best guess. We distinguish stable Roche lobe overflow (RLOF)
(where the behaviour of the mass loser is well understood) accompanied by the mass transfer (conservative or
non-conservative this is the question), mass accretion (always accompanied by momentum accretion causing a
spin-up of the mass gainer), common envelope evolution (very non-conservative: fact), the supernova (SN)
explosion of one of the binary components (the SN is in many cases asymmetrical: fact) which disrupts many
systems (fact), the spiral-in process of binaries with extreme mass ratio (some massive binaries survive
this process, survive two SNs and form a binary pulsar: fact), the merger process of the two components.
Due to the limited space it is unfeasible to discuss the facts and questions of these processes in detail. A
more extensive description was given in Vanbeveren et al. (1998a). In the present paper I prefer to propose
facts and questions related to a personal selection of observations of massive stars. 

\section{The surface chemistry of massive stars}

The surface layers of many OB supergiants and of some OB-dwarfs are nitrogen enriched (fact). Five processes
may be  responsible: rotational mixing combined with stellar wind mass loss, the 
RLOF process in interacting binaries where the surface layers of the mass loser but also of the
mass gainer may becomes N-enriched, the merger of two binary components due to a highly
non-conservative RLOF (common envelope phase) and last not least, the collision and merger process
due to N-body dynamics in young dense stellar systems. 

\medskip
\noindent Question: if RLOF has played an important
role during the progenitor evolution of WR+OB binaries and of high mass X-ray binaries (HMXBs), we
may expect that the OB companions in these systems are N-enriched. Has this been observed, in some
or in a majority of systems?

\section{The stellar wind mass loss rates}

The stellar wind mass loss rates during the core hydrogen burning (CHB) phase 
before the star eventually becomes a luminous blue variable (LBV) has been the subject of
detailed research during the last 3 decades. Here we focus on the LBV phase, the red
supergiant (RSG) phase and the Wolf-Rayet (WR) phase during core helium burning (CHeB).  

Hydrodynamic simulations (within its limitations) let us suspect that rotating stars (even those with
moderate rotational velocities) with a luminosity close enough to the Eddington value, will lose mass at
very high rate (Aerts et al., 2004 and references therein). LBVs are stars with $\gamma = L/L_{Edd}$ close
to 1. One of the most famous LBVs is $\eta$ Car, a star with $\gamma > 0.7$ and an observed average mass
loss rate  $\sim$ $10^{-3}$ M$_{\odot}$/yr (high + low state). Observations reveal that $\eta$ Car type
LBVs are hydrogen burning stars with a luminosity larger than the maximum luminosity of RSGs. Notice that
this max luminosity corresponds to the evolutionary track of a 40-50 M$_{\odot}$, in the Galaxy and in the
Magellanic Clouds. The interpretation could be the following: LBV-type mass loss is large enough
in order to prevent a star to expand and to become a RSG and detailed evolutionary calculations
show that a rate $\sim$ $10^{-3}$ M$_{\odot}$/yr is indeed sufficient in order to prevent the redward
evolution of a star with an initial mass $\ge$ 40 M$_{\odot}$. The consequence for binaries is then obvious:
RLOF does not happen (or its importance is significantly reduced) in case B/C and late case A binaries with
primary mass larger than 40-50 M$_{\odot}$ (the {\it LBV scenario of very massive stars} as it was
introduced in Vanbeveren, 1991). This LBV-type instability may be the reason why no or very few stars are
observed with a luminosity (with corresponding mass) larger than the luminosity of a 100-120 M$_{\odot}$
evolutionary track (an exception may be the Pistol Star, Figer et al., 1998). Evolutionary
calculations reveal that already on the zero age main sequence, a 150 M$_{\odot}$ has a $\gamma$ = 0.9, a
200 M$_{\odot}$ even has $\gamma$ = 0.96. This let us suspect that rotating stars with an initial mass $>$
120 M$_{\odot}$ may suffer from a very high stellar wind mass loss rate already at zero age. The
consequences are obvious: when a star is formed with a mass $>$ 120 M$_{\odot}$, it will very soon evolve
into a state where it is almost undistinguishable from a star whose mass was $\sim$ 120 M$_{\odot}$ on
the zero age main sequence.  

Notice that the process discussed above also affects in a critical way the outcome of N-body dynamical
computations of young dense stellar systems, and in particular the formation of intermediate mass black
holes (IMBH)  by runaway collision (Portegies Zwart et al., 2004). Simulations performed by Belkus et al.
(2005, see also the present proceedings) reveal that due to the action of an LBV type instability in
stars with a mass $>$ 120 M$_{\odot}$ formed by stellar collisions (remark that a collision implies spinning
up of the merger and this amplifies the LBV type instability and mass loss rate), the formation of
an IMBH by runaway collision in young dense stellar systems is less likely.  

Our RSG stellar wind mass loss knowledge was and is still very bad. The effect on stellar evolution has been
investigated by the Geneva group using a formalism proposed by de Jager et al. (1988), by the Brussels
group using a formalism  proposed by Vanbeveren et al. (1998a, b) and by the Padua group using a formalism
presented by Salasnich et al. (1999) which is quite similar to the one of Vanbeveren et al. These
3 formalisms have been critically discussed by Crowther (2001). Notice that the main evolutionary
difference between the Geneva treatment and the Brussels-Padua one concerns stars with an initial
mass $\le$ 25-30 M$_{\odot}$. When the RSG wind formalism proposed by the Brussels-Padua group is
implemented into a stellar evolutionary code, it follows that all galactic single stars with a mass between
$\sim$ 20 M$_{\odot}$ and the LBV-mass limit of 40-50 M$_{\odot}$ quoted above lose most of their
hydrogen rich layers during the RSG phase and become WR stars. The consequence for binaries: RLOF and thus
common envelope evolution does not happen in case C binaries with primary mass larger than $\sim$ 20
M$_{\odot}$ (the {\it RSG scenario} as it was introduced in Vanbeveren, 1996). Even more: using the RSG
formalism also for stars with initial mass $\le$ 20 M$_{\odot}$, it follows that stars with a mass as small
as 10 Mo may lose a significant amount of hydrogen rich layers (but not all) by RSG wind. Remark that
Vanbeveren et al. (1998) proposed a model for the binary $\upsilon$ Sgr where the effect of the RSG wind
is essential. Even more, it was concluded that without this RSG wind it is impossible to explain the
system.  

The masses of black hole (BH) remnants of stars with initial mass $\le$ 120 M$_{\odot}$  predicted by
stellar evolutionary computations depend on the effect of stellar wind mass loss on massive star evolution,
and more specifically on the mass loss during the CHeB-WR-phase. Before 1998, most of the massive star
evolutionary calculations used a WR-mass loss rate formalism which was based on theoretical interpretation
of WR spectra with atmosphere models that assume homogeneity of the stellar wind (Hamann, 1994). However,
two years later, Moffat (1996) and Hillier (1996) presented evidence that WR winds are inhomogeneous
implying that the real WR mass loss rates were smaller by at least a factor 2-3. In 1998, we were among the
first to perform and publish evolutionary computations of massive stars with such reduced WR mass loss rates
(Vanbeveren et al., 1998a, b, c). At that time, the evolutionary-referees were not always in favor, to
express it mildly. After 1998, observational evidence was growing that indeed WR winds are inhomogeneous
and that the rates are lower. Since 2000, everybody is using reduced WR-rates in their evolutionary code
(probably also our 1998-referees). A major consequence of lower WR mass loss rates is of course the final
stellar mass before core collapse. In our 1998 calculations, stars with a metallicty Z = 0.02 and
with an initial mass $\le$ 120 M$_{\odot}$ end their life with a mass $\le$ 20 M$_{\odot}$. When the WR
stellar wind mass loss rate is metallicity dependent (as predicted by the radiation driven wind theory,
Pauldrach et al., 1994 among many others), the pre-core collapse mass may be as large as 40-50 M$_{\odot}$
in small Z environments (like the SMC for example).

\section{Observations of massive close binaries (facts)}

We obviously have the numerous observations of individual massive systems, mainly in the Solar 
Neighborhood which can learn us a lot about the overall evolution of massive stars. Many of these
have been discussed in Vanbeveren et al. (1998).  

Less than 50\% of the WR stars in the Galaxy and the Magellanic Clouds seem to have an OB type companion
(Foelmi et al., 2003); the massive O-type binary frequency in open clusters in the Solar Neighborhood
ranges between 14\% (Trumpler 14) and 80\% (IC 1805) (Mermilliod, 2001); on average 33\% of the O-type and
early B-type stars in the Solar Neighborhood are primary of a binary with mass ratio secondary mass/primary
mass q $>$ 0.2 and orbital period P $<$ 100 days (Garmany et al., 1980; Vanbeveren et al., 1998a). But
accounting for statistical bias, the real massive close binary frequency may be significantly larger
(Halbwachs, 1987; Hogeveen, 1991; Mason et al., 2001; Van Rensbergen, 2001).

\section{The massive star population}

A massive star population is a mixture of unevolved binaries (no interaction yet), evolved 
binaries (interaction happened), single stars born as single, single stars who became single due
to binary evolution (disrupted binaries and mergers). In the case of clusters the population may
also be significantly affected by close encounter stellar dynamics. All this means that the binary
frequency on the zero age main sequence may be significantly larger than the observed binary frequency in a
population.  Accounting for the observed OB-type binary frequency discussed in the
previous section, it can be concluded that theoretical population predictions which do not account for the
effects of binaries may have an academic value but may be far from reality.  

The effects of binaries on the O-type and WR-type star population have been discussed frequently in
papers published by our group (for a review, see Vanbeveren et al., 1998b). Notice that since 1998, our
population prediction of binaries with a compact companion has been updated using the two-component
maximum-likelihood kick velocity distribution of Arzoumanian et al., 2002. The new frequencies are
higher compared to our 1998 values, but the differences are small enough that they do not change
the overall conclusions.

A post-RLOF primary of an interacting binary (that survived the RLOF of course) is always a hydrogen
deficient CHeB star. As a consequence, the answers on the following questions affect significantly
theoretical population synthesis predictions of WR binaries: what is the minimum mass (luminosity) of a
hydrogen deficient CHeB star to show up as a WR star? The WR components in observed WR binaries have a mass
$\ge$ 5-8 M$_{\odot}$. If this is a real minimum then it can be expected that, compared to the number of
WR+OB binaries, there are many more OB stars with a post-RLOF CHeB companion which does not show up as a WR
star. 

A minimum mass of hydrogen deficient CHeB stars means a minimum luminosity and it is
tempting to translate this into a minimum mass loss rate, i.e. 

\begin{itemize}
\item what must be the minimum mass loss rate of a hydrogen deficient CHeB star so that this star shows up
as a WR star in a binary with an OB-type companion ? 

\item if the mass loss rate is metallicity dependent as predicted by the radiatively driven wind theory,
could it be that the minimum mass of a WR star in a binary is larger in the Magellanic Clouds compared to
the value holding for the Galaxy ?
\end{itemize}   

\subsection{The early Be type stars}

(Facts) 20\% of the Galactic early B-type stars are Be-type. Be-stars are rapid rotators however 
a significant fraction of the {\it normal} early B-type stars have rotational velocities similar to
those of the Be-stars (Vanbeveren et al., 1998c) which means that not all rapid rotators are Be
stars.  There are no Be+B or Be+Be binaries known but many Be stars are binary mass gainers (Be
X-ray binaries, $\phi$-Persei types discussed in these proceedings by Douglas Gies).
 
\medskip
\noindent Question: how many Be stars are binary products i.e. how many Be stars are formed by binary mass
exchange or are binary mergers ?  
\medskip

To answer this question it may be crucial to remark that some Galactic
and Magellanic Cloud clusters contain a large population of Be stars. Six clusters with an age between 19
Myr and 25 Myr, studied by Mermilliod (1981) and by Grebel (1995), have an early Be/B0-B5 number ratio
between 0.1 and 0.4. If these fractions can be considered as facts, then the population synthesis
study of Van Bever and Vanbeveren (1997) predicts that at most 10-20\% of these Be stars can be
explained by binary evolution.

\subsection{The standard high mass X-ray binaries (HMXBs)}

The scenario for the formation of HMXBs proposed by Van den Heuvel and Heise (1972) has been 
confirmed frequently by detailed binary evolutionary calculations. We distinguish three X-ray
phases: 1. the OB star is well inside its critical Roche lobe and loses mass by stellar wind. The
X-rays are formed when the compact star accretes mass from the wind (wind fed systems); 2. The OB
star is at the beginning of its RLOF phase and mass transfer towards the compact star starts
gently (RLOF fed systems); 3. The optical star is a Be star and X-rays are emitted when the
compact star orbits inside the disk of the Be star (disk fed systems).  

In the massive binary evolutionary simulations performed with the Brussels code, we detected a possible
fourth phase: when the OB+compact companion binary survives the RLOF-spiral-in-common envelope phase and the
optical star is at the end of its RLOF, burning helium in its core, it transfers mass at a very
moderate rate similar as the rate at the beginning of RLOF. The star is overluminous with respect
to its mass, the surface layers are nitrogen rich and have a reduced surface hydrogen abundance (X
$\le$ 0.4). Possible candidates with an overluminous optical companion are Cen X-3 and SMC X-1. The
question here is how a binary can survive the spiral-in phase? Obviously, some binaries have to
survive because we observe double neutron stars. Theoretically the survival probability becomes
larger if one accounts in detail for the combined action of stellar winds and spiral-in. To
illustrate, when after the formation of the compact star, the binary period is large enough so
that an LBV-type or RSG-type stellar wind mass loss can start before the onset of the spiral-in,
the importance of the latter process can be reduced significantly. Our simulations with the wind
rates discussed in section 3 allow to conclude that it cannot be excluded that some HMXBs are
RLOF-fed systems where the optical star is a core helium burning star at the end of the RLOF/spiral-in.
 
Most of the supernova type Ib/c happen in binaries and all HMXBs with a neutron star companion are
expected to have experienced (and survived) such a supernova. Since a WC star is expected to be a
type Ic progenitor, evolutionary calculations predict that the SN shell may contain lots of carbon
and oxygen. When this WC star was a binary component and when the SN shell hits the OB companion
star, quite some C and O may be accreted by the latter and abundance anomalies may be expected.
Performing a detailed analysis of the CO abundanceÕs in the optical star of Cyg X-1 may be
interesting. An observed overabundance may be an indication that the black hole progenitor
experienced a supernova explosion as well, may be even a hypernova.

\section{The effect of rotation on massive star evolution}

Rotation implies rotational mixing in stellar interiors and it can enhance the stellar wind mass loss
compared to non-rotating stars. This enhancement may be important for stars that are close to the Eddington
limit (LBVs and very massive stars) and therefore rotation may affect indirectly their evolution.

The observed distribution of rotational velocities has been investigated by Vanbeveren et al.  (1998c) and
we illustrated that the majority of the early B-type stars and of the O-type stars are relatively slow
rotators, slow enough to conclude that rotational mixing only plays a moderate role during their evolution
(the effect is similar to the effect of moderate convective core overshooting). In the latter paper we
argued that due to the process of synchronization in binaries, accounting for the observed binary period
distribution, a majority of primaries in massive interacting binaries is expected to rotate slow enough so
that the effect of rotation on their overall evolution is moderate as well.  

The distribution has an extended tail towards very large rotational velocities, i.e. the distribution is
highly asymmetrical which means that in order to study the effect of stellar rotation on population
synthesis (the WR and the O type star population for example), it is NOT correct to use a set of
evolutionary tracks calculated with an average rotational velocity corresponding to the observed average.
This tail obviously demonstrates that there are stars which are rapid rotators. Binary mass
gainers, binary mergers and stellar collision products in young dense stellar environments are expected to
be rapid rotators and thus are expected to belong to the tail. The question however is
whether or not one can approximate their evolution with rotating single star models.  

Due to the dynamo effect, rotation generates magnetic fields (Spruit, 2002) which means that the
evolutionary effect of rotation cannot be studied separately from the effects of magnetic fields. This was
done only since recently (Maeder and Meynet, 2004; see also Norbert Langer in the present proceedings) and
(as could be expected) several of the stellar properties (size of the core, main sequence lifetime, tracks
in the HR diagram, surface abundanceÕs etc.) are closer to those of models without rotation than with
rotation only. Maeder and Meynet argued that since single star evolution with rotation only
explains the surface chemistry of the observed massive supergiant population, whereas single star
evolution with rotation and magnetic fields does not, magnetic fields must be unimportant. However this
argumentation is based on the assumption that most of the massive stars evolve as single stars do. In
section 2 I have listed the presently known processes which may be responsible for altered CNO abundances
in OB-stars and before an argumentation as the one above has any meaning, one has to consider all these
processes.

\section{A {\it UFO-scenario} for WR+OB binaries}

The formation of WR+OB binaries in young dense stellar systems may be quite different from the 
conventional binary evolutionary scenario as it was proposed by Van den Heuvel and Heise (1972).
Mass segregation in dense clusters happens on a timescale of a few million years which is
comparable to the evolutionary timescale of a massive star. Within the lifetime of a massive star,
close encounters may therefore happen very frequently. When we observe a WR+OB binary in a dense
cluster of stars, its progenitor evolution may be very hard to predict. In Brussels we started a
project to follow the dynamical evolution of starbursts. The first results are given by Belkus et
al. (2005)(see also Belkus et al. in the present proceedings). We combine the N-body integration technique
 with our detailed population number and spectral synthesis code of starbursts (Van Bever and Vanbeveren,
2000, 2003). Although we are still in the initial phase of this type of research, our simulations predict
the following unconventionally formed object scenario ({\it UFO-scenario}) of WR+OB binaries. After 4
million years the first WR stars are formed, either single or binary. Due to mass segregation, this happens
most likely when the star is in the starburst core. Dynamical interaction with another object becomes
probable, especially when the other object is a binary. In our simulations, we encountered a situation
where the WR star (a single WC-type with a mass = 10 M$_{\odot}$) encounters a 16  M$_{\odot}$ + 14
M$_{\odot}$ circularized binary with a period P = 6 days. A possible result of the encounter process is the
following: the two binary components merge and the 30 M$_{\odot}$ merger forms a binary with the WC star
with a period of $\sim$ 80 days and an eccentricity e = 0.3. This binary resembles very well the WR+OB
binary $\gamma$$^2$-Velorum but it is clear that conventional binary evolution has not played any role in
its formation.

\end{document}